\documentclass[reprint,superscriptaddress,aps,prb,twocolumn]{revtex4-1}
\usepackage{setspace}
\usepackage{amsmath}
\usepackage{graphicx}
\usepackage{amsfonts}
\usepackage{amssymb}
\usepackage{epstopdf} 
\usepackage{xcolor}
\usepackage{verbatim}
\usepackage{soul}
\usepackage{comment}
\usepackage{placeins}
\usepackage{textcomp}
\usepackage{booktabs}     

\usepackage{pdfpages} 
\usepackage{pgffor} 

\makeatletter
\AtBeginDocument{\let\LS@rot\@undefined}
\makeatother

\def\supplementfilename{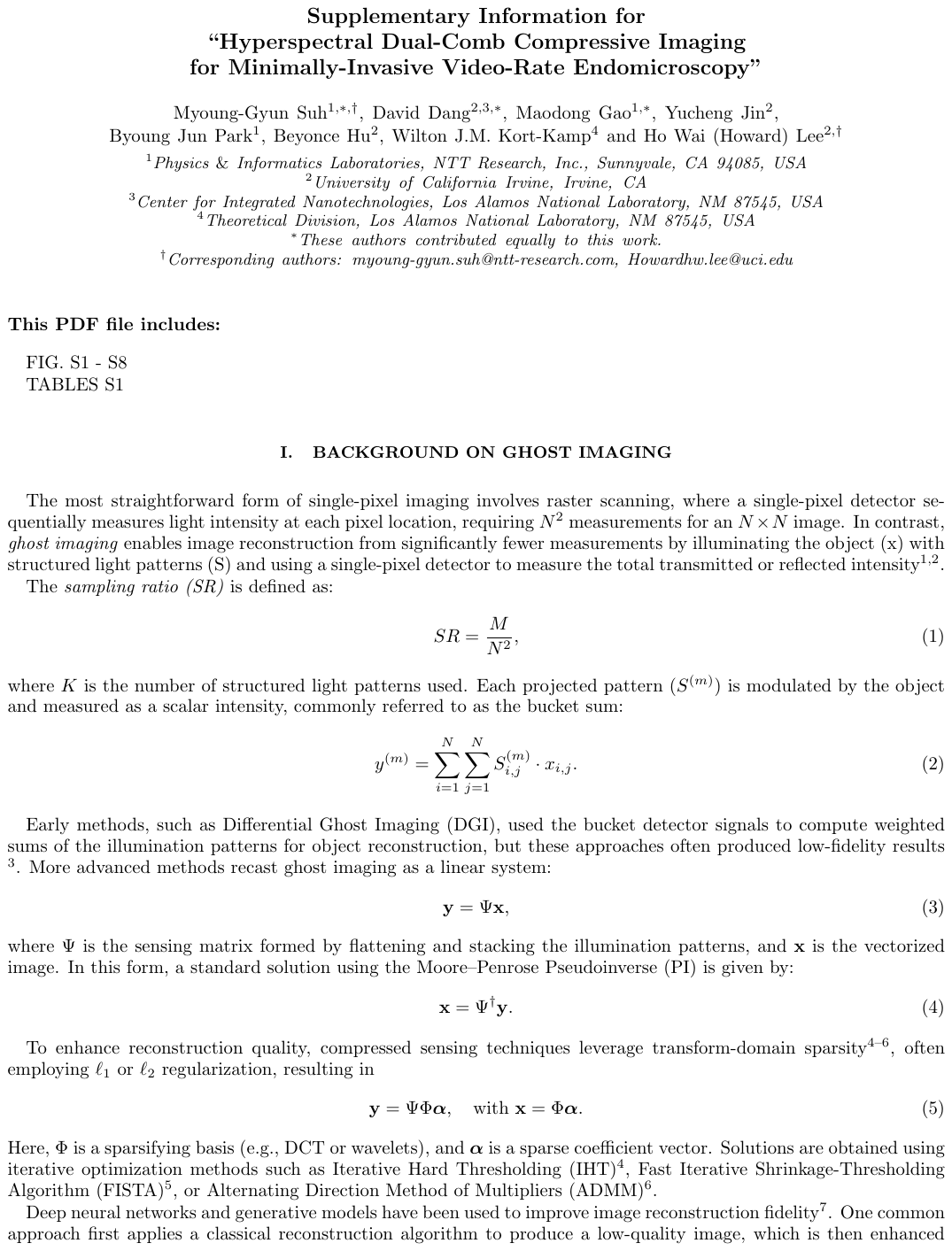}

\pdfximage{\supplementfilename}
\def\numbersupplementpages{\the\pdflastximagepages}

\newif\ifarXiv
\arXivtrue 

\usepackage[pagewise]{lineno}

\DeclareGraphicsExtensions{.pdf,.eps,.png,.jpg,.mps} 

\usepackage{float}

\begin{document}

\title{Hyperspectral Dual-Comb Compressive Imaging\\for Minimally-Invasive Video-Rate Endomicroscopy}




\author{%
Myoung-Gyun Suh$^{1,*,\dagger}$, David Dang$^{2,3,*}$, Maodong Gao$^{1,*}$, Yucheng Jin$^{2}$,\\ Byoung Jun Park$^{1}$, Beyonce Hu$^{2}$, Wilton J.M. Kort-Kamp$^{4}$
}
\author{Ho Wai (Howard) Lee$^{2,\dagger}$}\noaffiliation

\address{%
Physics $\&$ Informatics Laboratories, NTT Research, Inc., Sunnyvale, CA 94085, USA\\
$^2$University of California Irvine, Irvine, CA\\
$^3$Center for Integrated Nanotechnologies, Los Alamos National Laboratory, NM 87545, USA\\
$^4$Theoretical Division, Los Alamos National Laboratory, NM 87545, USA\\
$^*$These authors contributed equally to this work.\\
$^\dagger$Corresponding authors: myoung-gyun.suh@ntt-research.com, Howardhw.lee@uci.edu
}


\maketitle

\noindent{\textbf{
Endoscopic imaging\cite{berci2000history} is essential for real-time visualization of internal organs, yet conventional systems remain bulky, complex, and expensive due to their reliance on large, multi-element optical components. This limits their accessibility to delicate or constrained anatomical regions. Achieving real-time, high-resolution endomicroscopy using compact, low-cost hardware at the hundred-micron scale remains an unsolved challenge\cite{boese2022endoscopic,kiesslich2004confocal,urayama1996evaluation}. Optical fibers offer a promising route toward miniaturization by providing sub-millimeter-scale imaging channels; however, existing fiber-based methods typically rely on raster scanning or multicore bundles, which limit the resolution and imaging speed\cite{ren2022achromatic,porat2016widefield,lee2022confocal,lee2010scanning,choi2022flexible}. In this work, we overcome these limitations by integrating dual-comb interferometry\cite{coddington2016dual,bao2019imaging,vicentini2021dual} with compressive ghost imaging and advanced computational reconstruction\cite{katz2009compressive,padgett2017introduction}. Our technique, hyperspectral dual-comb compressive imaging, utilizes optical frequency combs\cite{fortier201920} to generate wavelength-multiplexed speckle patterns that are delivered through a single-core fiber and detected by a single-pixel photodetector. This parallel speckle illumination and detection enable snapshot compression and acquisition of image information using zero-dimensional hardware, completely eliminating the need for both spatial and spectral scanning. To decode these highly compressed signals, we develop a transformer-based deep learning model\cite{vaswani2017attention,ren2022ghost,chen2025large} capable of rapid, high-fidelity image reconstruction at extremely low sampling ratios. This approach significantly outperforms classical ghost imaging methods\cite{katz2009compressive,padgett2017introduction,blumensath2009iterative,beck2009fast,boyd2011distributed} in both speed and accuracy, achieving video-rate imaging with a dramatically simplified optical front-end. Our results represent a major advance toward minimally invasive, cost-effective endomicroscopy and provide a generalizable platform for optical sensing in applications where hardware constraints are critical.
}}

\begin{figure*}[hbtp]
  \centering
  \includegraphics[width=17cm]{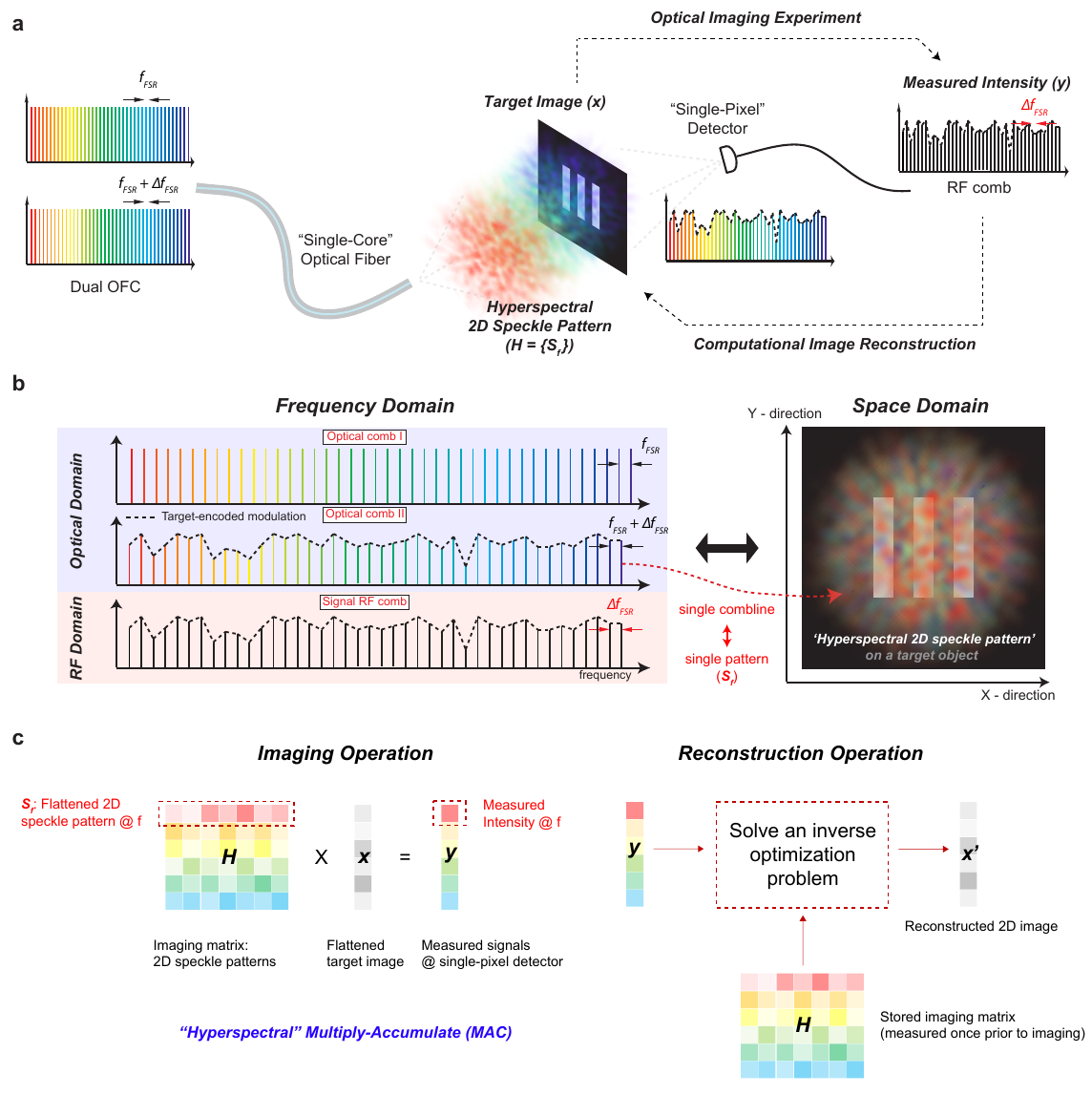}
\caption{
\textbf{Hyperspectral dual-comb compressive imaging.} \textbf{a,} Conceptual illustration of compressive ghost imaging with dual optical-frequency combs (OFCs). In the imaging experiment, a hyperspectral 2D speckle pattern ($H$), consisting of uncorrelated speckle patterns ($S_{f}$) at each comb line frequency, is generated at the fiber tip and mapped onto the target object $x$. The encoded speckle pattern ($H \times x$) is collected by a single-pixel detector, and the resulting dual-comb interference signal (``interferogram'') is recorded using an oscilloscope. A fast Fourier transform (FFT) converts this time-domain interferogram into a radio-frequency (RF) comb, yielding the bucket-sum signal ($y = H \times x$) used for computational image reconstruction. Thanks to comb-based wavelength-division multiplexing (WDM), dual-comb interferometry, and single-pixel compressive ghost imaging, the optical front-end hardware is reduced to a single-core fiber and a single-pixel detector, making the system well suited for compact applications such as endomicroscopy. \textbf{b,} In our approach, each optical comb line generates an uncorrelated speckle pattern, enabling parallel optical encoding of target object information without the need for spectral or mechanical scanning. The target-encoded optical modulation is then downconverted into the RF domain via dual-comb interferometry using two combs with slightly different line spacings, $f_{\text{FSR}}$ and $f_{\text{FSR}} + \Delta f_{\text{FSR}}$, allowing parallel, high-SNR bucket-sum measurements for fast, high-fidelity imaging. In our experiment, Comb I and Comb II are combined before illuminating the target. Alternatively, one comb can probe the target and be recombined with the other afterward; the latter configuration is shown for clarity. \textbf{c,} The imaging and computational reconstruction processes are illustrated in a simplified mathematical form. During imaging, the 3D hyperspectral hypercube $H$ (spanning frequency and 2D space) is multiplied by the 2D target object $x$, resulting in 1D bucket-sum data $y$ in the frequency domain. This operation is essentially a “hyperspectral” multiply-accumulate (MAC) process. The pre-recorded speckle patterns ($H$) and the measured bucket-sums ($y$) are then used to reconstruct the image by solving an inverse optimization problem.
}
\label{fig:fig1}
\end{figure*}

Conventional imaging systems rely on planar detector arrays to capture 2D spatial information, but their size and complexity limit applications requiring extreme miniaturization, such as in-vivo diagnostics. Flexible endomicroscopy, for example, demands sub-millimeter-scale form factors that rule out traditional cameras, filter wheels, and scanning mirrors. While endoscopes\cite{berci2000history} are standard for visualizing internal organs in procedures like gastrointestinal exams and bronchoscopies, their integrated optics and electronics make them bulky and often uncomfortable. Even with advances in compact CMOS sensors\cite{kim20181, omnivision}, overall system size still restricts resolution, especially in confined anatomical spaces\cite{boese2022endoscopic, urayama1996evaluation}. Optical fiber-based endomicroscopy\cite{kiesslich2004confocal,ren2022achromatic,porat2016widefield,turtaev2018high,gmitro1993confocal,jabbour2012confocal,flusberg2005fiber,lee2022confocal,lee2010scanning,choi2022flexible,sun2024lensless} offers a more compact solution, transmitting light through fibers as thin as 200–500 microns. Early systems using confocal or multiphoton fluorescence imaging achieved cellular-level resolution, while later advances introduced micro-electromechanical systems (MEMS) scanners, diffractive optical elements (DOEs), metalenses, and lensless or holographic methods\cite{lee2022confocal,kuschmierz2021ultra,ren2022achromatic,froch2023real,lee2010scanning,choi2022flexible,sun2024lensless}. However, these approaches still face challenges such as dispersion and dependence on mechanical scanning or fiber bundles, limiting speed and resolution.

In parallel with the development of endoscopic optical imaging technologies, we have also witnessed significant advances in optical frequency combs (OFCs) and their applications in imaging~\cite{bao2019imaging,hase2018scan,martin2020direct,sterczewski2019terahertz,giorgetta2024broadband} and spectroscopy~\cite{diddams2007molecular,Coddington2016DualCombReview,suh2016microresonator,suh2019searching}. OFCs are laser sources that emit a spectrum of equally spaced, mutually coherent narrow frequency lines—resembling the teeth of a comb—and can serve as broadband, high-precision optical sources for hyperspectral imaging\cite{martin2020direct,sterczewski2019terahertz,giorgetta2024broadband}. However, such implementations require complex wavelength filtering to read out the three-dimensional (2D spatial × 1D spectral) data hypercube. To enable snapshot detection of this 3D information, dual-comb interferometric readout has been applied at each pixel of 2D cameras in recent demonstrations of dual-comb hyperspectral imaging\cite{martin2020direct} and digital holography\cite{vicentini2021dual}. In addition, dual-comb imaging or microscopy using a single-core fiber and a single-pixel (or pinhole) detector—rather than a 2D camera—has been demonstrated by mapping each comb line directly to a single spatial pixel~\cite{Ideguchi2013SCAN,Picque2019CombImaging,bao2019imaging,hase2018scan}. Although such comb-based imaging approaches capture 2D spatial images rather than full 3D hyperspectral data, they offer advantages such as high frame rates and fill rates enabled by comb-based parallel processing\cite{bao2019imaging}. However, their spatial resolution is limited by the finite number of comb lines, and they typically require bulky dispersive or diffractive optical components, such as diffraction gratings or virtually imaged phased arrays (VIPAs).

Complementary to the pixel-by-pixel spatial mapping strategies discussed above, ghost imaging\cite{katz2009compressive,padgett2017introduction,li2021enhancing, liu2020photon, li2020compressive, lin2022ghost} introduces a fundamentally different approach that eliminates the need for pixelated sensors. Instead of direct spatial sampling, it reconstructs images by correlating known illumination patterns with total intensity measurements collected by a single-pixel (bucket) detector \cite{padgett2017introduction}. This technique enables computational image reconstruction through statistical analysis rather than direct spatial sampling. It has shown promise in low-light and scattering environments \cite{li2021enhancing, liu2020photon, li2020compressive, lin2022ghost}, and is especially suitable for endomicroscopy applications where multi-pixel detectors are impractical \cite{don2019introduction, kilic2022compressed}. Nonetheless, traditional ghost imaging suffers from slow acquisition speeds and limited fidelity due to sequential pattern projection and iterative reconstruction. 

In this work, we address these challenges by combining dual-comb interferometry with compressive ghost imaging to enable snapshot speckle imaging through a single-core fiber. By mapping comb lines to 2D speckle patterns and applying a dual-comb hyperspectral multiply-accumulate (MAC) “reduction” operation~\cite{latifpour2024hyperspectral}, our approach eliminates the need for spectral or mechanical scanning and bulky optics, enabling a truly zero-dimensional front-end imaging hardware: a single-core fiber for illumination and a single-pixel detector for signal acquisition. For image reconstruction, we develop a transformer-based deep learning model\cite{vaswani2017attention} that achieves real-time, high-fidelity image reconstruction even at sampling ratios as low as 0.3\%, overcoming the limitations of traditional methods\cite{blumensath2009iterative,beck2009fast,boyd2011distributed,liu2020imaging}. This hyperspectral dual-comb compressive imaging technique enables video-rate, high-quality endoscopic imaging with extremely compact hardware, making it well-suited for scenarios constrained by hardware size and complexity.


\begin{figure*}[hbtp]
\centering
\includegraphics[width=17cm]{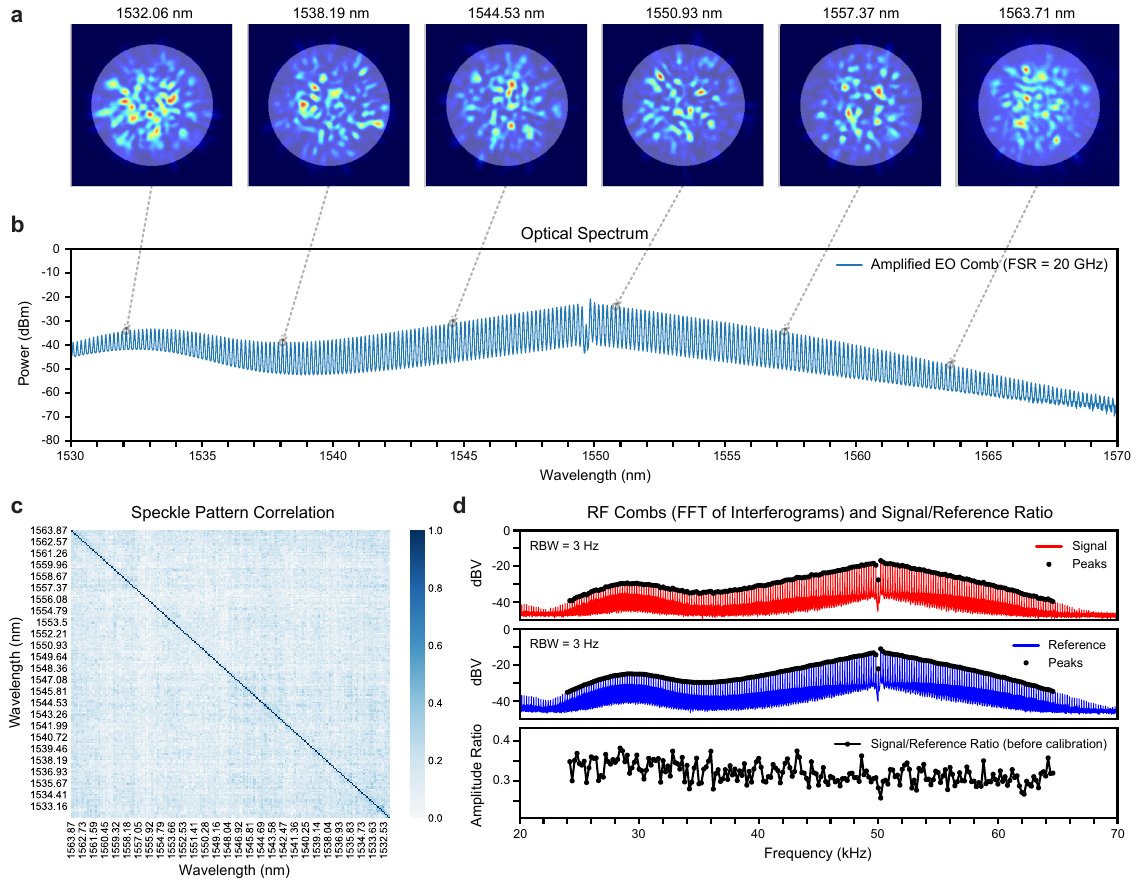}
\caption{
\textbf{Experimental details.} \textbf{a,} Speckle patterns measured at six different comb line frequencies. Each image contains $256 \times 256$ pixels and exhibits a circular speckle pattern centered within the frame. \textbf{b,} Typical optical spectrum of the electro-optically (EO) modulated comb with a 20~GHz free spectral range (FSR). The EO comb is generated from a 1550~nm continuous-wave (CW) fiber laser via resonant EO modulation and subsequently amplified by an erbium-doped fiber amplifier (EDFA) before being coupled into the experimental setup. \textbf{c,} Pearson correlation coefficients calculated between speckle patterns across the wavelength range from 1532~nm to 1564~nm. The correlation is computed within the bright circular region of each speckle pattern. The low off-diagonal values indicate that the speckle patterns are largely uncorrelated. Residual correlations are attributed primarily to the common background near the edges of the speckle patterns. \textbf{d,} Typical radio-frequency (RF) comb spectra generated from dual-comb interference signals, as measured by the signal and reference single-pixel detectors (upper and middle panels). The ratio of the comb peak amplitudes between the signal and reference RF combs is shown in the bottom panel. Differences between the signal and reference beam paths are calibrated using measurements performed with and without the target object. RBW: resolution bandwidth.
}
\label{fig:fig2}
\end{figure*}

Figure 1a illustrates the concept of our hyperspectral dual-comb compressive imaging method, which utilizes combline-to-speckle pattern mapping (Figure 1b). The method consists of two main steps: the optical imaging experiment and computational image reconstruction (Figure 1c).

In the optical imaging experiment, a set of speckle patterns $H$ = $\{ S_{f} \}$, consisting of uncorrelated 2D speckle patterns $S_{f}$ at different comb line frequencies $f$, is generated at the fiber tip using a dual-OFC source and projected onto the 2D target image $x$. The encoded intensity distribution ($H \times x$) is detected by a single-pixel photodetector. Because each comb source has a slightly different free spectral range (FSR), $f_{FSR}$ and $f_{FSR} + \Delta f_{FSR}$, the resulting time-varying voltage signal, containing the dual-comb interference (“interferogram”), is recorded with an oscilloscope. A fast Fourier transform (FFT) converts this time-domain signal into radiofrequency-domain bucket-sum data, enabling parallel and precise electronic detection of $y = H \times x$. To extract the target image information and suppress temporal noise, simultaneous dual-comb measurements are performed in both the signal path containing the target object and the reference path without the target (see Methods). At a high level, this information processing approach achieves data dimensionality reduction through a hyperspectral MAC operation\cite{latifpour2024hyperspectral}, as well as hardware dimensionality reduction through wavelength-division multiplexing (WDM) and dual-comb readout—enabling minimal front-end imaging hardware and significantly increasing imaging speed. The mutual coherence of the dual combs allows coherent averaging to improve the signal-to-noise ratio (SNR)~\cite{coddington2016dual}, which is another key advantage of the dual-comb method.

We first imaged static USAF 1951 target patterns using approximately 200 comb lines from each of the 20 GHz electro-optic (EO) combs spanning 1532--1564 nm (Figure 2b), with $\Delta f_{FSR} = 200$ Hz (see Methods). The speckle patterns generated at different comb-line frequencies exhibit distinct spatial distributions (Figure 2a), and Pearson correlation analysis confirms that the speckle patterns are largely uncorrelated with 20 GHz spacing (Figure 2c). Residual correlations are mainly attributed to common dark background components, especially near the pattern edges. Each comb line has a different power level, so the speckle patterns are normalized accordingly. Figure 2d shows the bucket-sum measurements obtained from the signal and reference RF combs (via FFT of the interferograms), along with their ratio, which contains the encoded image information and is used for image reconstruction. The same measurement is also performed without the object and is used to calibrate optical path imbalances between the signal and reference arms (see Methods).


For computational image reconstruction, we used the calibrated bucket-sum signals together with the pre-recorded speckle patterns. Figure 3a shows the bucket-sum signal after calibration, with the dual-comb experimental data in good agreement with the simulated prediction. The reconstructed image using the Moore–Penrose pseudoinverse (PI) algorithm clearly reveals the object, as shown in Figure 3b. However, the reconstruction exhibits significant background noise arising from the speckle patterns. As an aside, speckle patterns generated by different media (e.g., various MMFs) display different speckle grain sizes, corresponding to distinct narrowband spatial frequencies. To improve image resolution and fidelity, several approaches can be considered to generate speckle patterns with broader spatial frequency content. For example, the speckle pattern generator could be designed such that speckle patterns from different comb lines have different grain sizes, and thus different spatial frequencies. Additionally, as shown in Figure 3c, adjusting the beam spot size on the object can effectively modify the spatial frequency of the speckle patterns and enhance image resolution.

To further enhance image quality, we developed a transformer-based image reconstruction algorithm~\cite{ren2022ghost,chen2025large} that significantly reduces background noise. The model, which we call Ghost-GPT, compresses the speckle patterns into a latent space and concatenates them with the corresponding bucket measurements to form the input token sequence. This sequence is then processed through a stack of 12 transformer blocks, each comprising a multi-head self-attention (MHSA) mechanism and a two-layer feedforward network. The model outputs a reconstructed image with a resolution of 256 × 256 pixels. A detailed description of the model architecture and performance analysis is provided in the Supplementary Information.

\begingroup
  \setlength{\tabcolsep}{12pt}  
  \begin{table*}[hbtp]
    \caption{MSE/SSIM for Experimental Targets with Classical Algorithms and Ghost-GPT}
    \label{table:Experimental Metrics}
    \centering
    \begin{tabular}{lllll}
      \hline
      \cmidrule(r){1-2}
      Target     & DGI  & PI &  FISTA & Ghost-GPT (Ours)\\
      \hline
      Stripes & 0.231/0.030 & 0.140/0.028 & 0.072/0.057 & \bfseries 0.045/0.604 \\
      Number 2 & 0.204/0.036 & 0.138/0.028 & 0.084/0.064 & \bfseries 0.058/0.658 \\
      \hline
    \end{tabular}
  \end{table*}
\endgroup

\begin{figure*}[hbtp]
\centering
\includegraphics[width=17cm]{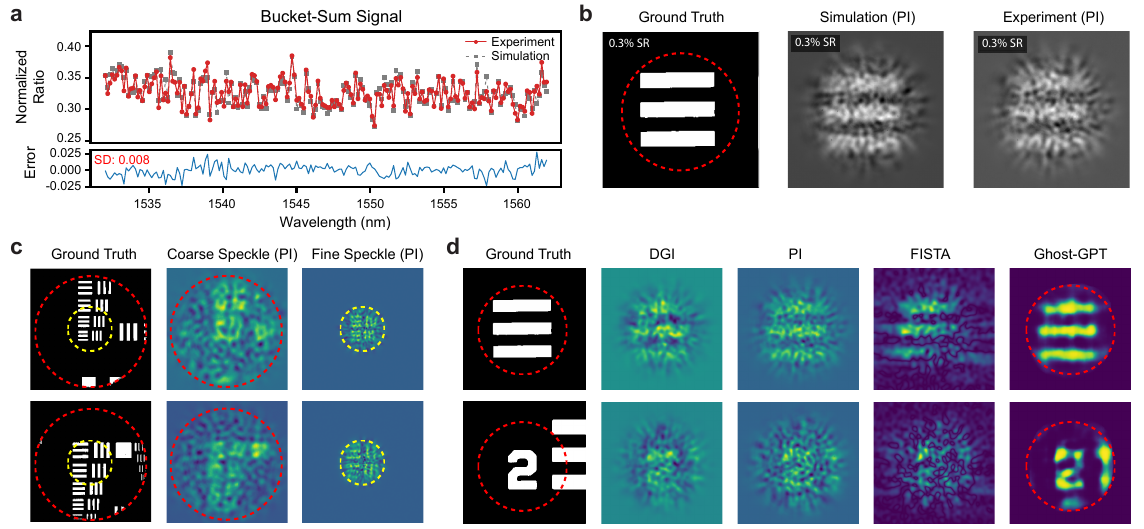} 
\caption{
\textbf{Image reconstruction of static targets.} 
\textbf{a,} The calibrated bucket-sum signal measured from the dual-comb experiment shows excellent agreement with the simulated bucket-sum signal, which is obtained by masking the ground-truth pattern onto the stored speckle pattern images. The standard deviation (SD) between the two signals is 0.008. \textbf{b,} Image reconstruction is performed using the Moore–Penrose Pseudoinverse (PI) algorithm at a sampling ratio (SR) of 0.3\%. Both the simulation (left) and experiment (right) successfully recover the ground-truth patterns, with residual background noise originating from the original speckle pattern. \textbf{c,} Image reconstruction of smaller patterns in group 2 (4.00 - 7.14 lp/mm). Fine details cannot be resolved when using a large illumination area with coarse speckle grains, but become resolvable with fine speckle grains when the beam is focused on a smaller area. Red and yellow dotted circles indicate the illuminated areas in each case. \textbf{d,} To improve reconstructed image quality, we developed a transformer-based deep learning model, Ghost-GPT. Reconstructions from experimental bucket-sum measurements using Ghost-GPT are compared with classical algorithms: Differential Ghost Imaging (DGI), Moore–Penrose Pseudoinverse (PI), and the Fast Iterative Shrinkage-Thresholding Algorithm (FISTA). (Top row): Striped lines correspond to group 0, element 4 (1.41 line pairs per millimeter). (Bottom Row): The number 2 corresponds to group 0, element 2 (1.12 lp/mm).
}
\label{fig:fig3}
\end{figure*}

Using the Ghost-GPT model, we were able to reconstruct high-fidelity images, achieving a Structural Similarity Index Measure (SSIM) greater than 0.6 and a Mean Squared Error (MSE) less than 0.05, despite a sampling ratio of only 0.3\% (see Table I and Figure 3d). The model effectively preserves fine structural details and clear object boundaries while suppressing background noise, highlighting its ability to recover meaningful image content from sparse measurements. We also compared our model against classical reconstruction algorithms and found that Ghost-GPT outperforms them in reconstructed image fidelity, while being approximately 430× faster than the Fast Iterative Shrinkage-Thresholding Algorithm (FISTA)\cite{beck2009fast}. The model’s rapid reconstruction speed of 8.39 ms enables real-time video-rate ghost imaging in optical fibers. Importantly, while the computational reconstruction speed can be further improved with more powerful computing hardware, the fundamental frame-rate limit is set by $\Delta f_{FSR}$ of the dual comb, which in our experiments ranges from a few hundred Hz to several kHz.


To demonstrate the high-speed imaging capability of our approach, we set $\Delta f_{FSR}$ = 3 kHz and imaged a target pattern moving at 2.4 mm/s (see Figure 4). For video-rate imaging at 60 Hz frame rate, the 1-second-long signal and reference interferograms (Figure 4b, middle panel) were divided into 60 time intervals. Each interval (Figure 4b, top panel) was Fourier-transformed to extract the bucket-sum signal. The Ghost-GPT model was then used to reconstruct images for each interval (frames 0–59), with five representative frames shown in the bottom panel of Figure 4b. At $\Delta f_{FSR}$ = 3 kHz, higher frame rates of up to 1 kHz were achievable, though with slight degradation in the SNR of the bucket-sum signals and in the quality of the reconstructed images (Figure 4c).


\begin{figure*}[hbtp]
  \centering
  \includegraphics[width=17cm]{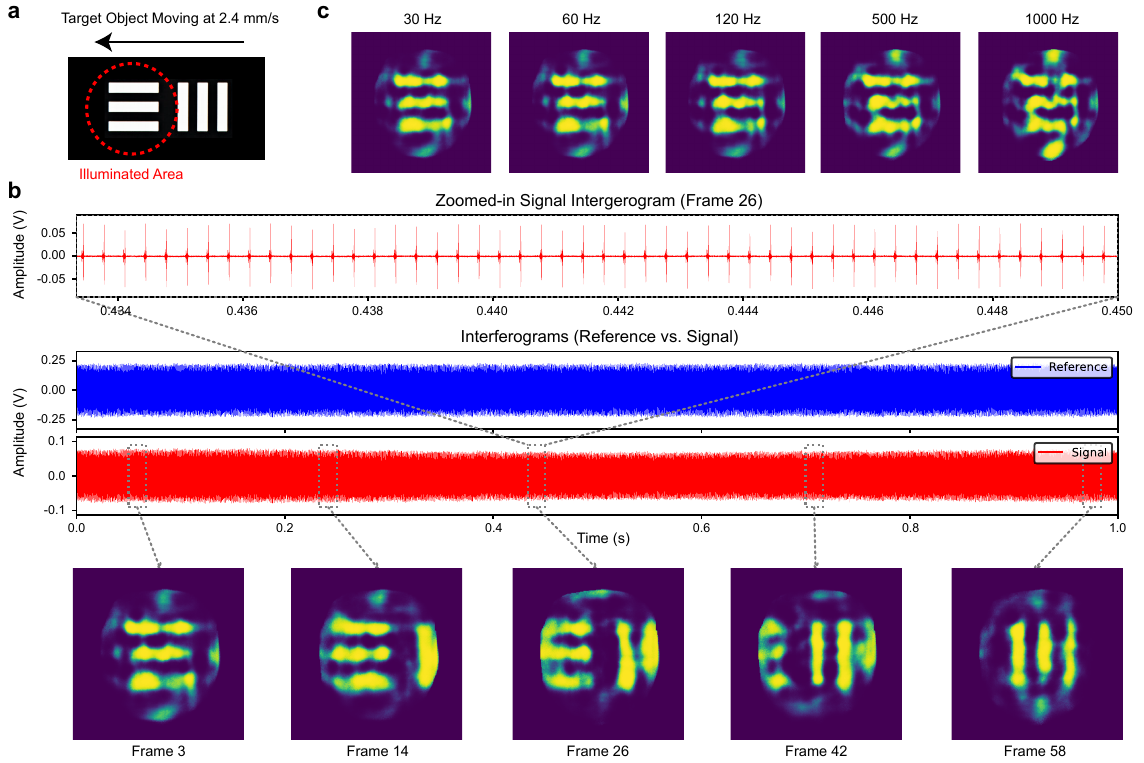}
\caption{
\textbf{Video-rate image reconstruction of a moving target.} 
\textbf{a,} A hyperspectral speckle pattern is projected onto a USAF1951 resolution target (Group 0, Element 4; 1.41 lp/mm), which is mounted on a motorized stage and translated horizontally at a speed of 2.4 mm/s (the maximum speed of the stage). \textbf{b,} Reference and signal interferograms of the moving object acquired over 1 second (middle panel), along with a zoomed-in view of the signal interferogram (top panel) used to reconstruct frame 26. The bottom panel displays five reconstructed images at 60 Hz frame rate, capturing the object’s motion. \textbf{c,} Reconstructed images of the object at t = 50 ms (corresponding to the start time of frame 3), obtained from the interferograms in (b), at five different frame rates (f = 30, 60, 120, 500, and 1000 Hz). Given the dual-comb repetition rate difference of 3 kHz, each image is reconstructed from 3000/f electrical pulses. The results demonstrate that imaging a much faster moving target is feasible.
}
\label{fig:fig4}
\end{figure*}



The demonstrated image quality and frame rate can be further improved with future system developments. In our current experiments, we selected comb spacings of 20 GHz or 25 GHz, considering the full-width at half-maximum (FWHM) of speckle pattern correlation and the line-by-line spectral filtering resolution of the Waveshaper used to resolve individual speckle patterns. In principle, sub-GHz FWHM speckle correlation is achievable~\cite{redding2014high,facchin2024determining}, and speckle pattern acquisition at such narrow comb spacings could be realized using a tunable CW laser calibrated against the comb~\cite{baumann2014comb}. Combined with spectral broadening of OFCs, this approach would allow a significantly larger number of speckle patterns to be used in image reconstruction, thereby improving both image fidelity and resolution\cite{chen2025large}.

For instance, with a 1 GHz comb spacing across a 100 nm spectral span near 1550 nm (yielding approximately 12,500 comb lines), a $\Delta f_{FSR}$ of 10 kHz, and a sampling ratio (SR) of 1\%, high-definition (HD) resolution video-rate imaging could be achieved with an effective fill rate of 12 gigapixels per second (\text{Fill rate} = $\Delta f_{FSR}/M \times N \times 1/SR$). Here, the frame rate ($\Delta f_{FSR}/M$) is determined by the free spectral range difference divided by the number of averaging periods $M$ required to enhance SNR. Currently, our system’s frame rate is primarily limited by the photodetector bandwidth ($f_{BW}$) and the number $N$ of radiofrequency comb lines, governed by the condition $N \times \Delta f_{FSR} < f_{BW}$. In principle, both the frame rate and the sampling ratio can be flexibly tuned depending on application requirements---balancing frame rate against image fidelity within the detector bandwidth limit.

The Transformer-based deep-learning model used for image enhancement can be further optimized, but improvements in the raw data quality---such as using low-noise photodetectors or cameras---would significantly enhance the quality of reconstructed images. System performance could also benefit from operating within wavelength ranges detectable by high-speed, high-sensitivity CMOS image sensors and photodetectors. Squeezed dual-comb sources offer an additional opportunity to improve performance by boosting the SNR and thereby reducing the required averaging time~\cite{herman2025squeezed}. Although our current demonstration is limited to 2D spatial imaging, the same system is capable of operating in spectroscopic imaging or spectroscopy-only modes, depending on the application~\cite{vicentini2021dual}. Furthermore, combining our spatial-domain compressive imaging with time-domain compressive sensing using time-programmable frequency combs~\cite{caldwell2022time, giorgetta2024free} would be an intriguing direction for future research, enabling simultaneous compression of the object's four-dimensional hyperspectral spatio-temporal information across both spatial and temporal domains.

In this work, we present an experimental demonstration of hyperspectral dual-comb compressive imaging, achieving video-rate imaging with minimal front-end hardware---a single-core optical fiber and a single-pixel photodetector. Such minimal hardware requirements could enable not only compact but also cost-effective, single-use endomicroscopic probes for medical applications. Furthermore, we show that image quality can be significantly enhanced by the Transformer-based reconstruction model proposed in this study, effectively minimizing information loss during the compression process. While several technical challenges still remain for practical applications---such as developing compact and reliable speckle pattern (or structured light) generators\cite{shin2016single}, efficiently collecting light on the photodetector, and achieving gray-scale imaging\cite{chen2025large}---our proof-of-concept demonstration shows that video-rate, high-fidelity endoscopic imaging is possible even with minimally invasive, zero-dimensional hardware configurations. This approach, powered by comb-based parallel optical processing and deep-learning-based computational ghost imaging, holds significant promise for a range of sensing applications constrained by hardware size and complexity.

\begin{footnotesize}

\section*{Methods}
\noindent \textbf{Experimental Setup}
For the dual-comb source, we use two electro-optic (EO) combs with slightly different free spectral ranges (FSRs). A continuous-wave (CW) fiber laser at $1550\,\mathrm{nm}$ is first amplified and split into two beams via a 50/50 fiber coupler. Each beam is frequency-shifted using an acousto-optic frequency shifter, introducing a center frequency offset $\Delta f_{\text{center}}$. These beams are then independently modulated by resonant EO modulators driven at $f_{\text{FSR}}$ and $f_{\text{FSR}} + \Delta f_{\text{FSR}}$, respectively, where $f_{\text{FSR}} = 20\,\mathrm{GHz}$ and $\Delta f_{\text{FSR}} = 200\,\mathrm{Hz}$ for static object imaging experiment (or $f_{\text{FSR}} = 25\,\mathrm{GHz}$ and $\Delta f_{\text{FSR}} = 3\,\mathrm{kHz}$ for moving targets). The resulting EO combs are recombined using another 50/50 fiber coupler and amplified to compensate for insertion loss of the downstream system.

Before entering the free-space imaging setup, a Waveshaper is placed in the system to optionally filter individual comb lines. For bucket-sum measurements, the full dual-comb spectrum is transmitted through the Waveshaper. Downstream, the 5-meter multimode fiber (MMF) used in the experiments has a core diameter of 200 microns and supports hundreds of spatial modes, resulting in distinct speckle patterns when light is outcoupled into the free-space setup. The generated speckle patterns are collimated and passed through the target object, a negative USAF 1951 resolution target mounted on a motorized translation stage. A 50/50 beam splitter divides the beam into signal and reference paths before the target, allowing simultaneous acquisition of signal and reference interferograms and thereby suppressing common-mode temporal noise. Both the reference and signal bucket-sums are acquired using free-space InGaAs photodetectors with a 510 kHz bandwidth. The reference path is in free space, although it could alternatively be implemented in fiber using a 50/50 fiber coupler, which may be more practical for real-world applications. Additionally, the imaging experiment can be conducted in reflective mode rather than the current transmissive configuration. The speckle pattern set ($H$) without the object is separately recorded using a 2D InGaAs camera by selecting individual comb lines with the Waveshaper, either before or after imaging. It is worth noting that the MMF was fixed to the optical table, and the resulting speckle patterns remained stable enough to support multiple imaging experiments over several hours. (Details of the experimental setup are provided in the Supplementary Information.)

\medskip

\end{footnotesize}

\bibliography{main}


\section*{Acknowledgements}

\section*{Author Contributions}



\noindent \textbf{\large Data Availability} \\ The data that support the plots within this paper and other findings of this study are available from the corresponding authors upon reasonable request.

\vbox{}
\noindent \textbf{\large Code Availability} \\ The code that support the plots within this paper and other findings of this study are available from the corresponding authors upon reasonable request.

\newpage


\ifarXiv
  

\section*{Supplementary material}

# Supplementary Information for
## "Hyperspectral Dual-Comb Compressive Imaging for Minimally-Invasive Video-Rate Endomicroscopy"

Myoung-Gyun Suh[1,*,†], David Dang[2,3,*], Maodong Gao[1,*], Yucheng Jin[2],
Byoung Jun Park[1], Beyonce Hu[2], Wilton J.M. Kort-Kamp[4] and Ho Wai (Howard) Lee[2,†]

[1] Physics & Informatics Laboratories, NTT Research, Inc., Sunnyvale, CA 94085, USA
[2] University of California Irvine, Irvine, CA
[3] Center for Integrated Nanotechnologies, Los Alamos National Laboratory, NM 87545, USA
[4] Theoretical Division, Los Alamos National Laboratory, NM 87545, USA
[*] These authors contributed equally to this work.
[†] Corresponding authors: myoung-gyun.suh@ntt-research.com, Howardhw.lee@uci.edu

**This PDF file includes:**

FIG. S1 - S8
TABLES S1

## I. BACKGROUND ON GHOST IMAGING

The most straightforward form of single-pixel imaging involves raster scanning, where a single-pixel detector sequentially measures light intensity at each pixel location, requiring $N^2$ measurements for an $N \times N$ image. In contrast, *ghost imaging* enables image reconstruction from significantly fewer measurements by illuminating the object (x) with structured light patterns (S) and using a single-pixel detector to measure the total transmitted or reflected intensity[1,2].

The *sampling ratio (SR)* is defined as:

$$SR = \frac{M}{N^2},$$ (1)

where $K$ is the number of structured light patterns used. Each projected pattern ($S^{(m)}$) is modulated by the object and measured as a scalar intensity, commonly referred to as the bucket sum:

$$y^{(m)} = \sum_{i=1}^{N}\sum_{j=1}^{N} S_{i,j}^{(m)} \cdot x_{i,j}.$$ (2)

Early methods, such as Differential Ghost Imaging (DGI), used the bucket detector signals to compute weighted sums of the illumination patterns for object reconstruction, but these approaches often produced low-fidelity results [3]. More advanced methods recast ghost imaging as a linear system:

$$\mathbf{y} = \Psi\mathbf{x},$$ (3)

where $\Psi$ is the sensing matrix formed by flattening and stacking the illumination patterns, and $\mathbf{x}$ is the vectorized image. In this form, a standard solution using the Moore–Penrose Pseudoinverse (PI) is given by:

$$\mathbf{x} = \Psi^\dagger\mathbf{y}.$$ (4)

To enhance reconstruction quality, compressed sensing techniques leverage transform-domain sparsity[4–6], often employing $\ell_1$ or $\ell_2$ regularization, resulting in

$$\mathbf{y} = \Psi\Phi\boldsymbol{\alpha}, \quad \text{with } \mathbf{x} = \Phi\boldsymbol{\alpha}.$$ (5)

Here, $\Phi$ is a sparsifying basis (e.g., DCT or wavelets), and $\boldsymbol{\alpha}$ is a sparse coefficient vector. Solutions are obtained using iterative optimization methods such as Iterative Hard Thresholding (IHT)[4], Fast Iterative Shrinkage-Thresholding Algorithm (FISTA)[5], or Alternating Direction Method of Multipliers (ADMM)[6].

Deep neural networks and generative models have been used to improve image reconstruction fidelity[7]. One common approach first applies a classical reconstruction algorithm to produce a low-quality image, which is then enhanced

using a trained neural network—such as a CNN, U-Net, or, more recently, a diffusion model[8,9,10]. This method benefits from the ability to incorporate learned priors for denoising and super-resolution. However, its performance depends on the quality as well as speed of the initial reconstruction and requires large datasets of paired examples. Another approach uses an end-to-end strategy by embedding elements of the ghost imaging process directly into the neural network architecture[11,12]. This method is typically faster, as it bypasses classical reconstruction algorithms entirely. However, prior studies have been constrained to low-resolution images (e.g., 28×28) and relied on binary mask patterns. Complementing these challenges on the algorithmic side, hardware constraints remain a significant bottleneck; however, dual optical frequency combs offer a promising solution.

Variable Definitions:

- $S^{(m)} \in \mathbb{R}^{N \times N}$: The $m^{\text{th}}$ structured illumination pattern,

- $x \in \mathbb{R}^{N \times N}$: The unknown image to be reconstructed,

- $y^{(m)}$: Total detected intensity for the $m^{\text{th}}$ pattern,

- $\mathbf{y} \in \mathbb{R}^{M \times 1}$: Vector of all scalar intensity measurements,

- $\Psi \in \mathbb{R}^{M \times N^2}$: Sensing matrix composed of flattened patterns,

- $\mathbf{x} \in \mathbb{R}^{N^2 \times 1}$: Flattened version of the image $x$,

- $\Phi \in \mathbb{R}^{N^2 \times N^2}$: Sparsifying basis (e.g., DCT, wavelets),

- $\boldsymbol{\alpha} \in \mathbb{R}^{N^2 \times 1}$: Sparse coefficients in the transform domain.

## II. DUAL ELECTRO-OPTIC COMB SOURCE AND IMAGING SETUP

To generate the dual electro-optic (EO) comb source, a continuous-wave (CW) fiber laser operating at 1550 nm is first amplified and split into two paths using a 50/50 fiber coupler. Before the splitter, a Fiber Bragg Grating (FBG) filter is used to suppress amplified spontaneous emission (ASE) noise. Each path is polarization controlled and undergoes a frequency shift via an acousto-optic frequency shifter, introducing a center frequency offset $\Delta f_{\text{center}}$. The beams are then modulated independently using resonant EO modulators (OptoComb WTEC-02-25) driven at $f_{\text{FSR}}$ and $f_{\text{FSR}} + \Delta f_{\text{FSR}}$, where $f_{\text{FSR}} = 20\,\text{GHz}$ and $\Delta f_{\text{FSR}} = 200\,\text{Hz}$ for static imaging, or $f_{\text{FSR}} = 25\,\text{GHz}$ and $\Delta f_{\text{FSR}} = 3\,\text{kHz}$ for dynamic targets. The resulting EO combs are recombined through a second 50/50 fiber coupler and re-amplified to offset downstream insertion losses.

A Waveshaper (Coherent WS-01000B) is positioned before the free-space imaging setup to optionally filter individual comb lines. For bucket-sum acquisition, the full dual-comb spectrum is transmitted. A 5-meter-long multimode fiber (MMF) with a 200 μm core—supporting hundreds of spatial modes—produces speckle patterns upon outcoupling into free space. These speckles are collimated and projected onto a transmissive negative USAF 1951 resolution target mounted on a motorized translation stage.

Prior to the target, a 50/50 beam splitter separates the beam into signal and reference paths, enabling simultaneous acquisition of interferograms from both arms and reducing common-mode temporal noise. Both signal and reference bucket signals are recorded using free-space InGaAs photodetectors (Thorlabs PDA50B2) with a 510 kHz bandwidth and a 5 mm diameter active area. Since the actual focused beam spot has a diameter smaller than 1 mm, photodetectors with a smaller active area and higher bandwidth could be used in principle, especially when detecting a large number of comb lines or operating with a larger $\Delta f_{\text{FSR}}$. While the reference arm is implemented in free space here, a fiber-based version using a 50/50 fiber coupler is feasible and may be more practical for real-world systems. Reflective-mode imaging is also possible as an alternative to the current transmissive setup.

The speckle basis set ($H$)—recorded without the target—is acquired using a 2D InGaAs camera (Pembroke Instruments SenS-320V-ST) by selecting individual comb lines via the Waveshaper, either before or after imaging. The MMF was fixed to the optical table, ensuring speckle pattern stability over several hours, allowing for repeatable imaging experiments. In different imaging experiments, we used 150 to 200 comb lines across the optical C-band.

(a)

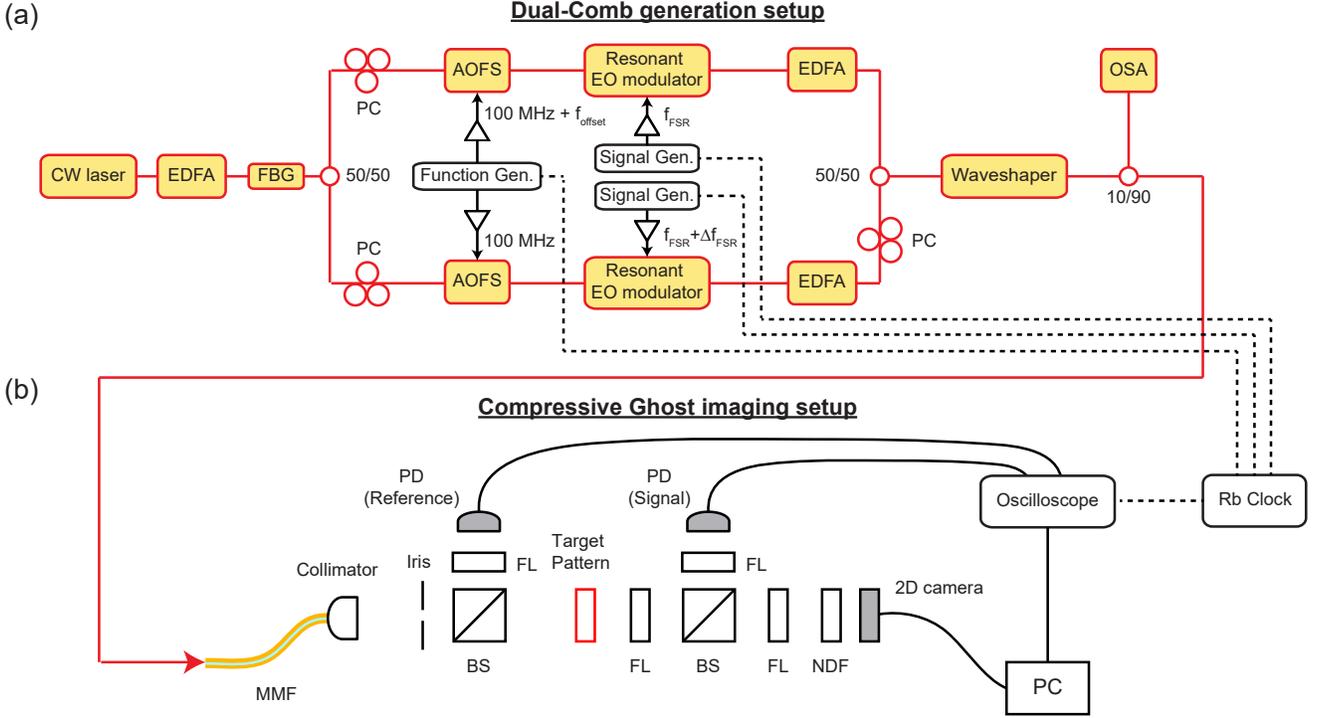

**FIG. S1: Experimental Setup** (a) Dual-Comb generation setup. (b) Free-space compressive Ghost imaging setup. For improved SNR, a function generator, two signal generators, and an oscilloscope were referenced to an external timing reference (Rb clock). EDFA: Erbium Doped Fiber Amplifier, FBG: Fiber Bragg Grating, AOFS: Acousto-Optic Frequency Shifter, SMF: Single Mode Fiber, MMF: Multi Mode Fiber, OSA: Optical Spectrum Analyzer, PD: Photodetector, FL: Focusing Lens, BS: Beam Splitter, NDF: Neutral Density Filter.

## III. GHOST-GPT: TRANSFORMER-BASED IMAGE RECONSTRUCTION MODEL

### A. Model Architecture

Transformers are a powerful deep learning architecture originally introduced for natural language processing (NLP) tasks but have since found applications in various domains[13], including computer vision[14], speech processing[15], and scientific data analysis[16]. They leverage a mechanism called self-attention to model long-range dependencies and capture contextual relationships within sequences[17]. Unlike traditional recurrent or convolutional networks, transformers process entire sequences simultaneously, making them highly efficient for parallel computation. Their ability to learn complex patterns from large datasets has made them the backbone of state-of-the-art models like Bidirectional Encoder Representations from Transformers (BERT), Generative Pre-training Transformer (GPT)[18], and Vision Transformers (ViTs)[19][20].

In typical ViTs, the image is broken up into equally sized patches, which serve as the ViTs' token. In order to adapt this methodology to dual-comb ghost imaging, we propose concatenating the flattened illumination patterns that make the sensing matrix, $\Psi$, with the bucket value to form the token for our model.

*a. Model Architecture.* We introduce **Optical Ghost-GPT** (or simply **Ghost-GPT**), a transformer-based model designed for structured image reconstruction from ghost imaging measurements. The model leverages stacked self-attention mechanisms and residual feedforward blocks to model long-range dependencies across the contextual input.

*b. Input Embedding.* The model receives two primary inputs in each token: a flattened image of the speckle pattern $\Psi^m \in \mathbb{R}^N$, where $N = 256 \times 256$, and bucket sum value $\mathbf{y^m}$. The image vector is projected via a learnable linear transformation to a latent embedding of dimension $d_k - 1$, and afterwards, the latent representation of the image vector and its corresponding bucket sum are concatenated. To encode positional structure, we add learned positional embeddings to each token in the sequence:

$$\mathbf{z}_i = \mathbf{e}_i + \mathbf{p}_i, \quad i = 1, \ldots, C,$$

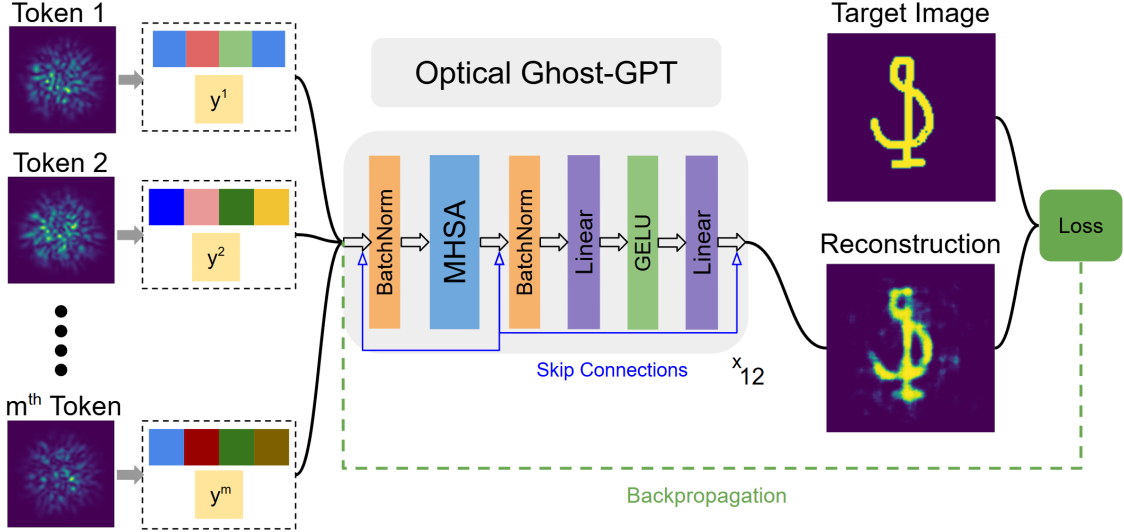

**FIG. S2: Schematic of Optical Ghost-GPT**. The speckle patterns are first compressed into a latent space and then concatenated with the corresponding bucket measurements to form the input token sequence.

where $\mathbf{e}_i$ is the input token embedding, $\mathbf{p}_i$ is the corresponding positional embedding, and C=250 is the context size determined by the theorectical maximum number of RF combines in our setup.

*c. Transformer Blocks.* The architecture contains a stack of $L = 12$ transformer blocks, each composed of a multi-head self-attention (MHSA) mechanism and a two-layer feedforward network. The attention mechanism employs $H$ number of heads, computed as:

$$\text{Attention}(Q, K, V) = \text{Softmax}\left(\frac{QK^\top}{\sqrt{d_k}}\right)V,$$

where $Q$, $K$, and $V$ are linear projections of the input sequence, and $d_k$ is the size of the latent space embedding. We also apply a dropout mask with a value of 0.1 to ensure robustness of the model and allow the model to generalize to missing shots and buckets. Each block includes residual connections and a batch normalization operation placed before and after the MHSA layer, defined as:

$$\text{BatchNorm}(\mathbf{x}) = \frac{\mathbf{x} - \mu}{\sqrt{\sigma^2 + \varepsilon}}, \quad \mu = \mathbb{E}[\mathbf{x}], \quad \sigma^2 = \text{Var}[\mathbf{x}].$$

The feedforward network consists of two linear transformations with a gaussian error linear unit (GELU) activation; A final transformer block is appended after the main stack to further refine the sequence representation.

$$\text{FFN}(\mathbf{x}) = \text{Linear}_2\left(\text{GELU}(\text{Linear}_1(\mathbf{x}))\right).$$

*d. Output Projection.* The final token representations are projected to $\mathbb{R}^{16}$ via a linear layer and the output is flattened to a tensor of size $(C \times 16)$ and passed through a final linear layer to reconstruct the original image vector in $\mathbb{R}^{256 \times 256}$. A sigmoid activation is applied to constrain the output to the range $[0, 1]$, consistent with normalized image intensities:

$$\hat{\mathbf{x}} = \sigma\left(\text{Linear}_{\text{final}}(\text{Flatten}(\mathbf{z}))\right).$$

*e. Network Training.* We first obtain a set of speckle patterns from the 2-D camera during the calibration phase of the experiment. (We emphasize that the 2-D camera is only needed to obtain the initial speckle patterns and can be removed during imaging). For this experiment, we obtained 188 speckle patterns, which are then used to generate synthetic buckets sums via a convolution between the digitized speckle pattern and images from the MNIST and OMNIglot datasets. We form our labeled dataset of synthetic bucket sums as the x-label and its corresponding target images as the y-label (See Section III D for dataset generation, train/test split). In training, Ghost-GPT predicts the target image, given the speckle pattern and bucket sums as the input. We use mean squared error as our loss function

and an AdamW optimizer with a learning rate of 0.0003 and a weight decay of 0.1.

## B. Ghost-GPT Reconstruction Results

In this section, we examine the reconstruction results in simulation using our experimental speckle pattern results. Figure S3 shows a series of reconstruction compared with their true images.

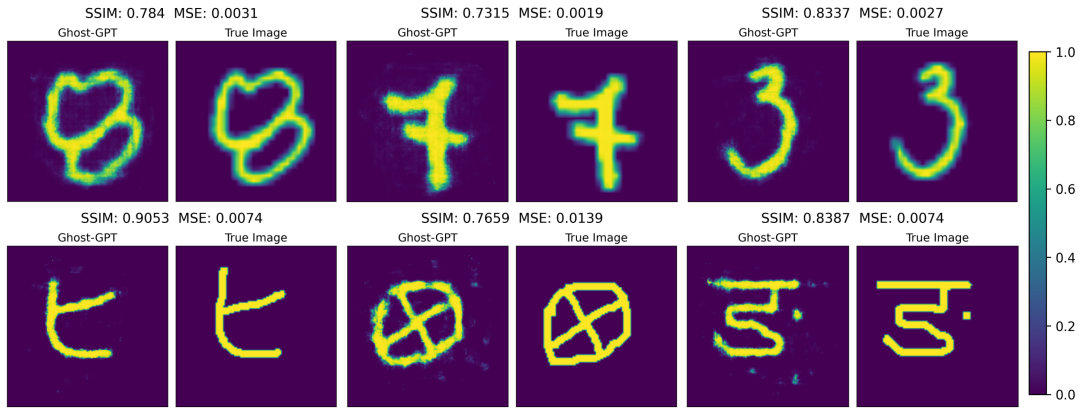

**FIG. S3:** Reconstructions from Ghost-GPT versus the true image with the Structural Similarity Index Measure (SSIM) and Mean Squared Error (MSE) displayed above.

We emphasize that we achieve very high SSIM (greater than 0.7) and low MSE (less than 0.02) values despite having a sampling ratio of only 0.29%. We observe that the model is able to preserve fine structural details and clear object boundaries, highlighting the model's ability to recover meaningful image content from minimal measurements. However, the reconstructions can exhibit variations in intensity across the object, especially in images with long, thin structures or uniform intensity profiles. These artifacts likely stem from the inherent non-uniformity of the speckle patterns generated by the multi-mode fiber. This issue could potentially be mitigated by employing a more uniform speckle distribution or by including a smoothness term, such as total variational loss, in the loss function.

We also compared our model against classical reconstruction algorithms on 256 images from our validation dataset. (256 images were chosen due to the long reconstruction times associated with the iterative FISTA algorithm). Table S1 compares the MSE and SSIM of previously discussed reconstruction algorithms- as expected the simpler reconstruction algorithms such as Differential Ghost Imaging and the Moore-Penrose Psuedo Inverse perform worse than compressed sensing methods like FISTA. However, Ghost-GPT outperforms these classical algorithms giving an average **MSE of 0.007** and **SSIM of 0.825**, while being approximately **430x** faster than FISTA. (We set $\epsilon = 50$ after performing a hyperparameter sweep and choosing the best performing SSIM). The extremely fast reconstruction speed of **8.39 ms** enables **real-time, video frame-rate ghost imaging in optical fibers**. Importantly, while the computational reconstruction speed can be further improved with better computing hardware, the fundamental limit of image reconstruction is set by the repetition rate difference of the dual-comb, which is typically a few hundred Hz to several kHz in our experiments. With a larger repetition rate difference, a much higher frame rate is possible. However, this requires high-bandwidth photodetection, and the trade-off between the sampling ratio and frame rate must be considered in accordance with the Nyquist condition.

**TABLE S1:** Classical Algorithms vs Ghost-GPT

| Algorithm | MSE | SSIM | Computational Time (ms) |
|---|---|---|---|
| Differential Ghost Imaging | $0.184 \pm 0.037$ | $0.042 \pm 0.022$ | $15.7 \pm 0.138$ |
| Moore–Penrose Pseudo-Inverse | $0.093 \pm 0.027$ | $0.055 \pm 0.027$ | $5640 \pm 52.1$ |
| FISTA (200 iters, $\epsilon$=50) | $0.045 \pm 0.017$ | $0.092 \pm 0.019$ | $3680 \pm 125$ |
| **Ghost-GPT (Ours)** | $\mathbf{0.007 \pm 0.010}$ | $\mathbf{0.825 \pm 0.092}$ | $\mathbf{8.39 \pm 0.090}$ |

### C.   Robustness of Ghost-GPT in Experimental Imaging

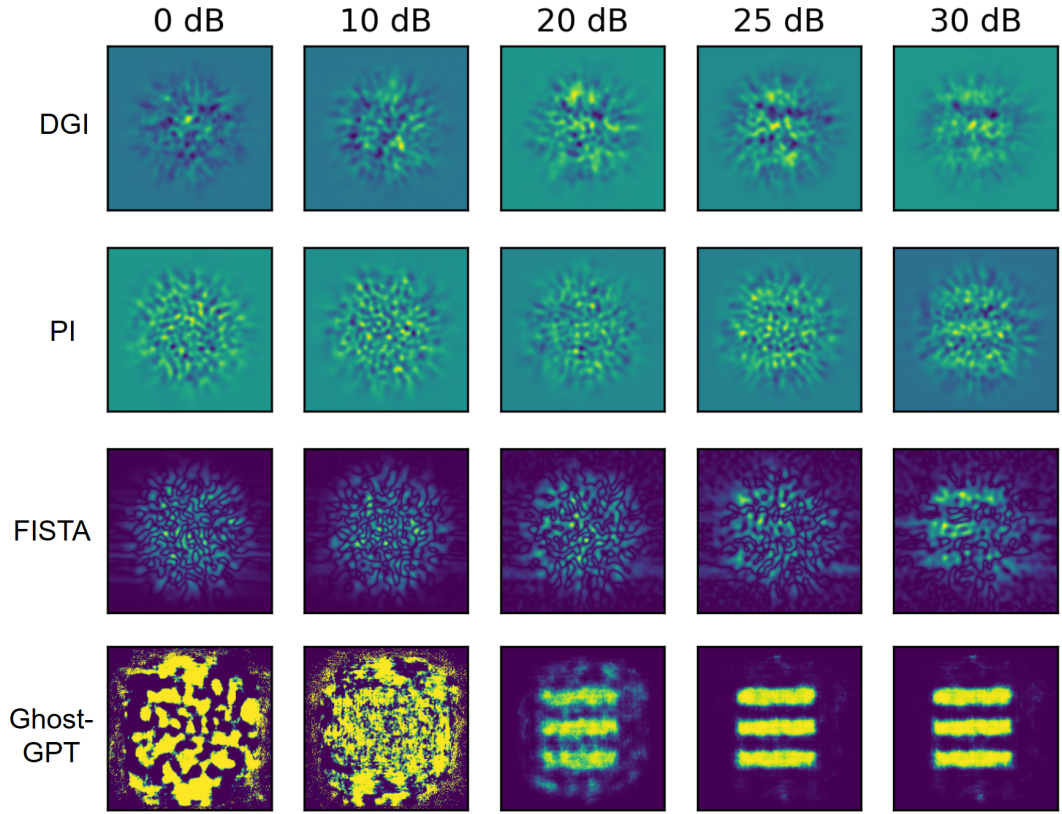

**FIG. S4:** Simulated reconstructions of the USAF striped targets with artificial noise added to the buckets. The signal to noise ratio in dB is shown label at the top with the corresponding reconstruction algorithms on the left-hand side.

To examine the effect of noise in our experimental measurement, we performed a signal-to-noise ratio analysis by adding varying amounts of noise to simulated buckets and gauged the quality of image reconstructions (See FIG. S4). The artificial noise added to the buckets were calculated by sampling points from a gaussian with mean of zero and standard deviation of one and scaling it inversely proportional to the value of the SNR (See Section III F for more information about the SNR calculation). Ghost-GPT demonstrates strong robustness to noise, outperforming classical reconstruction algorithms, with recognizable reconstructions achievable at SNR levels above 20 dB for Ghost-GPT. (See FIG. S6 for plots of the calculated MSE and SSIM values vs SNR). Based on the image reconstruction quality of both the classical reconstruction algorithm and Ghost-GPT in the SNR analysis, our experimental bucket measurements are approximately equivalent to 25-30 dB.

### D.   Dataset Generation, Train and Test Split

Since the bucket signal is effectively a convolution between the object and each speckle pattern, it is computationally efficient to synthetically generate labeled training data by pairing known objects with their corresponding bucket sums. For this, we use open-source image datasets and simulate the measurement process. Our dataset includes 19,280 from OMNIglot and 19,280 images from MNIST with a train/val split of 33,000/5,560 respectively. All images are resized to $256 \times 256$ to match the resolution of the recorded speckle patterns. We use a batch size of 32 for the training dataset and a batch size of 64 for the validation dataset.

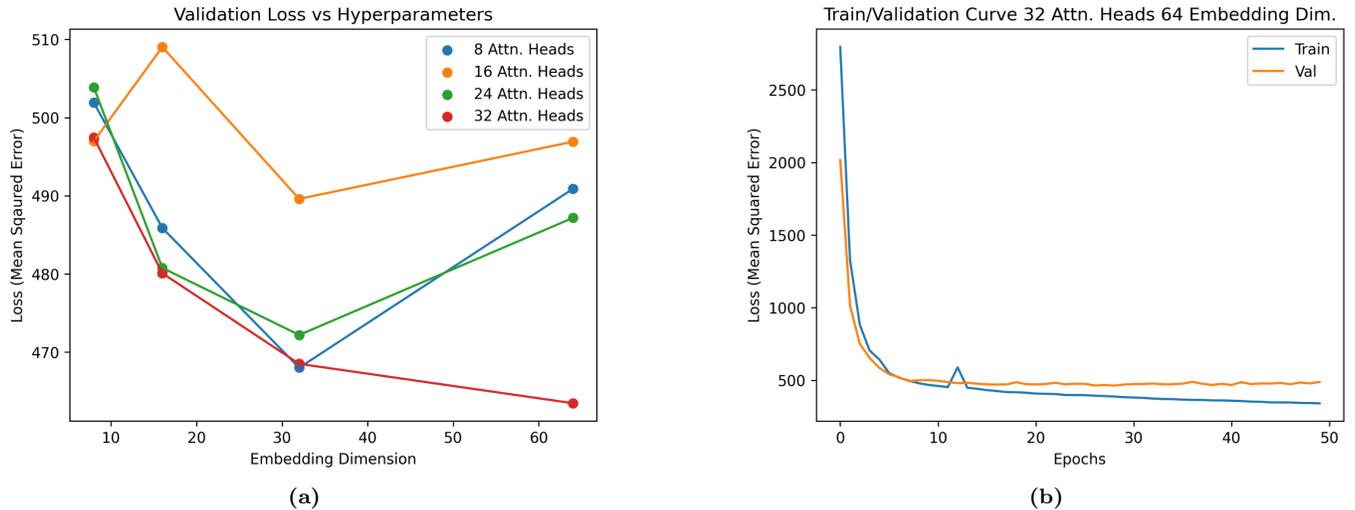

(a)                                                    (b)

**FIG. S5:** The lowest loss value on the validation dataset as a function of the number of attention heads and embedding dimensions (a) comparison with pattern embedding (b) log scale loss values of models with and without pattern embedding

### E. Hyperparameter Analysis

For hyperparameter tuning, we performed a simple parameter sweep by changing the number of attention heads at values of 8, 16, 24, and 32 as well as adjusting the embedding dimension at values of 1 (no pattern embedding), 8, 16, 32, and 64. As previously stated, we use mean squared error as our loss function and an AdamW optimizer with a learning rate 184 of 0.0003 and a weight decay of 0.1. In our loss function, we add the error in each pixel and then average over the batch size.

Over 50 epochs, the model took approximately 2 hours to train on a Linux workstation, using an AMD Ryzen Threadripper 3990X 64-Core Processor and a single NVIDIA A6000 GPU. Our largest model (32 attention heads, embedding dimension 64) contained $\sim$ 269 million trainable parameters and demonstrates convergence afters 10 epochs (see FIG S5b) - displaying no characteristics of overfitting. FIG. S5a shows that Ghost-GPT generally performs better as we increase the size of the model- indicating that with larger embedding dimensions the model is able to understand and model short/long range dependencies between the speckle patterns more effectively.

### F. Different Algorithms vs SNR

The SNR was calculated via the following equation:

$$\mathbf{y}_{noisy} = \mathcal{N}(0,1) * \sigma + \mathbf{y}, \quad \sigma = \frac{\mathbb{E}[\mathbf{y}]}{\mathrm{SNR}^{10}}$$

We sample points from a gaussian of mean 0 and std 1, which is scaled by $\sigma$, the standard deviation. $\sigma$ is calculated the by averaging over the collection of bucket sum and then dividing by the SNR value raised to the $10^{\mathrm{th}}$ power. The true bucket sum, $\mathbf{y}$, is added to the previously calculated quantity to create the noisy bucket, $\mathbf{y}_{\mathrm{noisy}}$. Previous works in ghost imaging formulated their SNR analysis using these exact definitions.

FIG. S6 plots the MSE and SSIM of the reconstructed images compared with classical reconstruction algorithms and Ghost-GPT.

### G. Resolution Analysis

In ghost imaging, resolution is fundamentally linked to the spatial characteristics of the light patterns used to probe the object—specifically, the grain size of these patterns. Grain size refers to the typical scale or correlation length of the intensity fluctuations in the illumination patterns (often speckle patterns). This grain size determines the finest detail that can be distinguished in the reconstructed image.

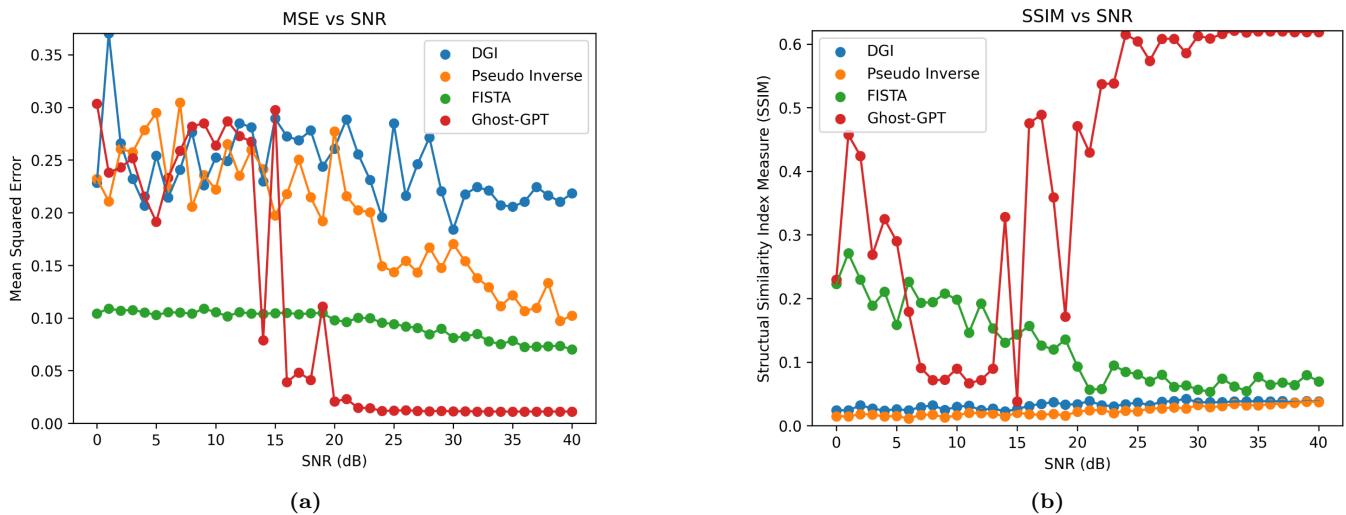

**FIG. S6:** SNR ratio in the buckets versus (a) MSE and (b) SSIM of the USAF target

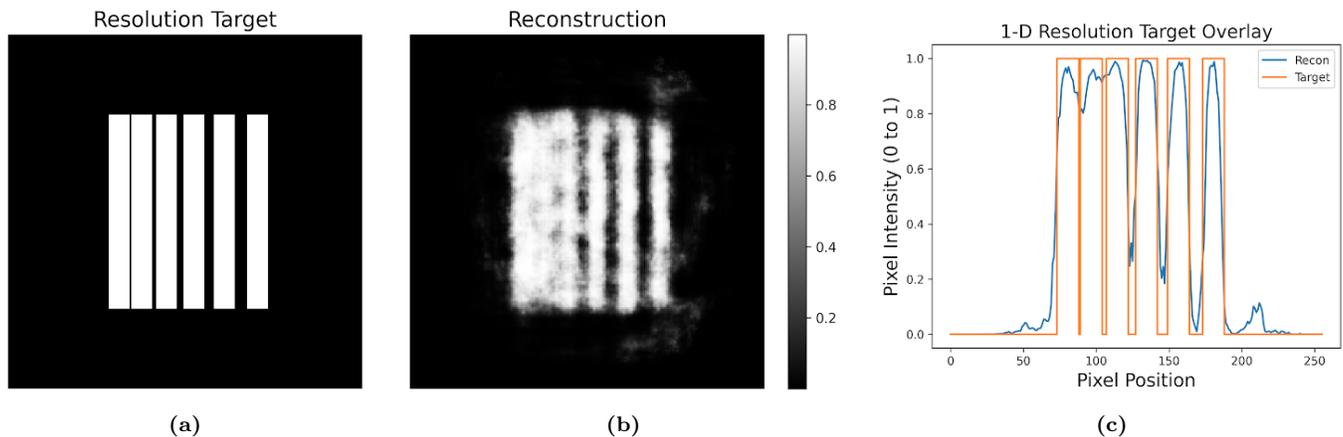

**FIG. S7:** Simulated resolution testing of Ghost-GPT. (a) True resolution target with increasing pixel gaps (b) Ghost-GPT reconstruction results (c) Line profile comparing the ground truth and model output, obtained by scanning along a horizontal cross-section midway through the images.

In this section, we evaluated the smallest resolvable feature size achievable by Ghost-GPT in simulation using our current experimental speckle pattern set (See FIG. S7. A digital resolution target was generated with 15-pixel-wide bars and varying gap sizes (1, 3, 5, 7, 9 pixels). Ghost-GPT successfully resolved features with a 5-pixel gap. In our experimental setup, the Air Force resolution target's 23-pixel feature between a pair of stripes corresponds to a physical size of 0.355 mm, indicating that Ghost-GPT achieves a resolution of approximately 0.077 mm. At this resolution, our ghost imaging model is well-suited for visualizing a broad range of biological structures, including tissue architecture such as blood vessels, skin layers, and muscle fibers. It also enables observation of developmental forms in small organisms like embryos and larvae, as well as structural features in plant tissues such as roots and leaves.

### H. Experimental Hyperparameter Tuning

Due to the effects of noise in the experimental data, the reconstructed images have visual artifacts and degraded image quality, such as the presence of rough edges and artificial spots. As a result, we tuned our model further by adjusting the mean squared error (MSE) loss function to also include a total variational (TV) term to help improve

the experimental image quality. The resulting equation is defined as:

$$Loss = MSE + \beta * TV \tag{6}$$

The addition of total variational loss in the model helps to smooth artifacts in the image at the cost of overall image sharpness. We weighed the TV by a factor, $\beta$, and performed a hyperparameter sweep between 0.1 and 0.5 in steps of 0.1 to find the right balance between the two terms in the loss function (FIG S8 and Table S2). The values of $\beta$ was determined by evaluating our model on our validation dataset with various amounts of gaussian noise added to the simulated bucket sum, with a SNR between 15 to 25 dB. We found that a $\beta$ value of 0.4 was the most optimal value for the experimental data with the highest average SSIM value.

Additionally, we found that cropping and down-sampling the 256×256 speckle patterns also improved image fidelity. Specifically, we removed low-intensity regions and down-sampled the speckle patterns to 64×64. This reduction in size allowed us to set the `embedding_dim` to a much larger size of 129 without hitting memory or runtime constraints. The underlying intuition is that the model can better learn the relationship between speckle patterns and bucket sums when the input is more compact and focused. These improvements to Ghost-GPT and its corresponding results are shown in the main manuscript.

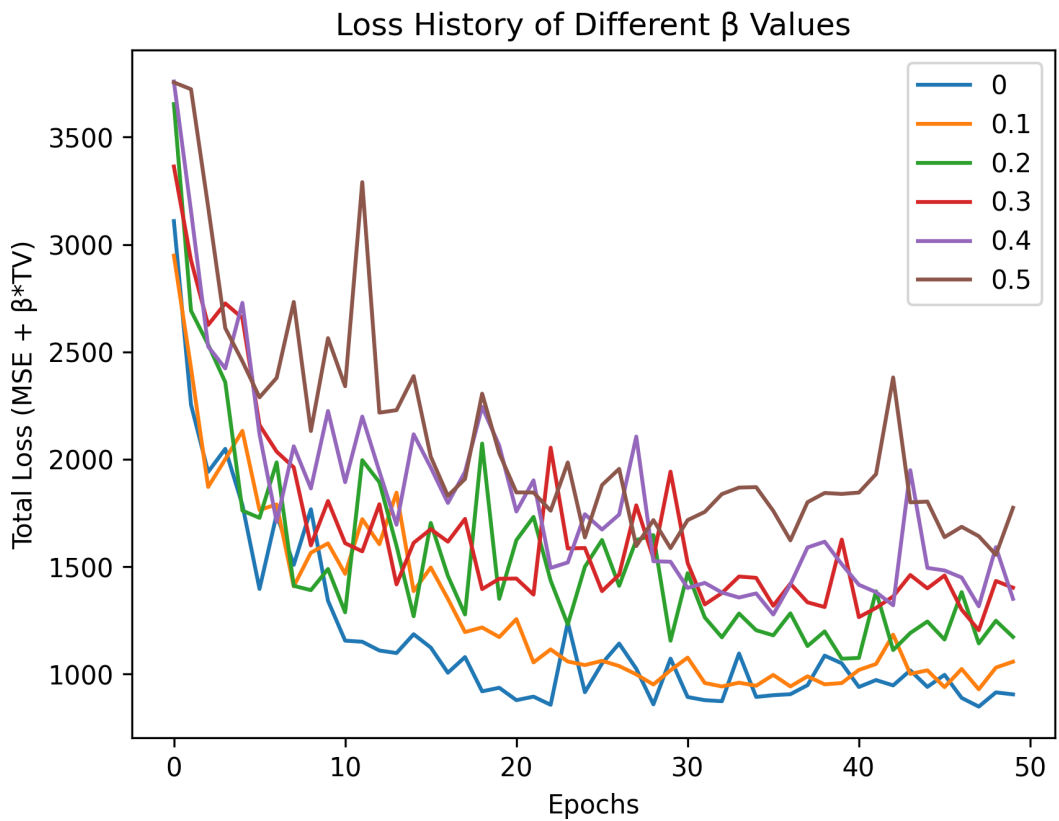

**FIG. S8:** Loss history with the total variational loss term, which is weighted by a scalar value $\beta$. The value of $\beta$ is shown by the different colors within the plot. Convergence of the network occurs within 50 epochs.

**TABLE S2:** Average SSIM on Noisy Validation Dataset

| TV weight ($\beta$) | 0 | 0.1 | 0.2 | 0.3 | 0.4 | 0.5 |
|---|---|---|---|---|---|---|
| SSIM | 0.74 ± 0.063 | 0.77 ± 0.070 | 0.75 ± 0.076 | 0.73 ± 0.145 | **0.80 ± 0.066** | 0.75 ± 0.077 |

# I. Limitations

While Ghost-GPT demonstrates high reconstruction fidelity, computational efficiency, and robustness to noise, there are several limitations with our current work. First, the model performance is contingent on the quality and representativeness of the calibration speckle patterns; any deviations in experimental conditions may lead to distribution shift, potentially reducing reconstruction accuracy. Second, although the model generalizes well to synthetic and experimental data within the resolution constraints of our setup, it has not been benchmarked on more complex, naturalistic scenes or dynamic objects beyond the training domain. Additionally, while the use of synthetic training data from MNIST and Omniglot enables rapid prototyping, the domain gap between these datasets and real biological or clinical targets may necessitate fine-tuning with domain-specific data. Third, our current framework is only comparing our model against classical algorithms.

In this work, we focused our comparison on classical ghost imaging reconstruction algorithms such as Differential Ghost Imaging, the Moore-Penrose Pseudoinverse, and FISTA. However, further benchmarking against advanced generative or super-resolution models—such as diffusion models, adversarial networks, or U-Nets —could offer insight into alternative priors and reconstruction strategies (i.e. more complex loss functions) that may enhance fidelity or reduce artifacts. We emphasize that this work represents an early proof-of-principle for dual-comb ghost imaging using a transformer-based architecture and should be interpreted as a foundation for future improvements. Continued development will include more sophisticated training data, model architectures, and experimental protocols to fully exploit the potential of this novel imaging paradigm.


\fi

\end{document}